\documentclass{an}
\usepackage{graphicx}
\usepackage{times}
\Volume{00}              % issue in current year, format {NN}
\Year{0000}              % format: {YYYY}
\Month{00}               % format: {M} or {MM} (number), month of print version
\Pagespan{000}{000}      % format: {start page}{last page}

\begin{document}
\lhead[\thepage]{A.N. Author: Title}
\rhead[Astron. Nachr./AN~{\bf XXX} (200X) X]{\thepage}
\headnote{Astron. Nachr./AN {\bf 32X} (200X) X, XXX--XXX}
\def\aap{Astron. \& Astroph. }
\def\pn{\par\noindent}
\def\chandra{{\it Chandra}}
\def\xmm{XMM--{\it Newton}}
\def\cgs{erg cm$^{-2}$ s$^{-1}$}
\def\gsimeq{\hbox{\raise0.5ex\hbox{$>\lower1.06ex\hbox{$\kern-1.07em{\sim}$}$}}} % maggiore uguale circa
\def\lsimeq{\hbox{\raise0.5ex\hbox{$<\lower1.06ex\hbox{$\kern-1.07em{\sim}$}$}}}

\title{Hard X--ray observations of Extremely Red Objects}

\author{Marcella Brusa} 
%\and A. Cimatti\inst{3}
%\and E. Daddi\inst{4}
%\and A. Comastri\inst{2}}
\institute{
Dipartimento di Astronomia, Universit\`a di Bologna, via Ranzani 1, 
I-40127 Bologna, Italy}
%\and 
%INAF -- Osservatorio Astronomico di Bologna, via Ranzani 1, I-40127
%Bologna, Italy}
%\and 
%INAF -- Osservatorio Astronomico di Bologna, via Ranzani 1, I-40127
%Bologna, Italy
%\and 
%INAF -- Osservatorio Astronomico di Bologna, via Ranzani 1, I-40127
%Bologna, Italy}
\date{Received {date will be inserted by the editor}; 
accepted {date will be inserted by the editor}} 

% ABSTRACT 
%######################################################################
\abstract{Extremely Red Objects (EROs, R$-$K$>$5) constitute a
heterogenous class of extragalactic sources including high redshift
elliptical galaxies, dusty star--forming systems and heavily obscured
AGNs. Hard X-ray observations provide an unique and powerful tool to uncover
obscured nuclear or star--forming activity.
We present the results of XMM--{\it Newton} observations
of the largest sample of EROs available to date
(about 450 objects over a contiguous area of 700 arcmin$^2$). Five of the 46
hard X--ray selected sources brighter than 3$\times 10^{-15}$ cgs in
the 2--10 keV band, are associated with EROs.
All of the X-ray detected EROs show rather extreme X--ray--to--optical flux
ratios, suggesting the presence of highly obscured AGN activity.
We also report on the X--ray stacking analysis of spectroscopically identified
EROs in the {\it Chandra} Deep Field South. 
}

\correspondence{brusa@bo.astro.it}

\maketitle

\section{Introduction}
%#######################################################
Extremely Red Objects (EROs, $R-K>5$)
show the bulk of the emission in the near-infrared band and are
associated with extremely faint optical counterparts that have so far
limited the identification process. The first available
optical--infrared spectra (e.g., Cimatti et al. 1998; Smail et
al. 1999; Pierre et al. 2001) and the recent results from the 
{\tt K20 survey} (Cimatti et al. 2002) indicate that 
EROs counterparts lie at high redshifts. 
In particular, their colors are consistent
with those of early--type passively evolving galaxies, dusty
star--forming systems and Active Galactic Nuclei (AGN) reddened by
dust and gas in the redshift range $z=0.8\div$ 3. 
The relative fraction of these different subclasses among the ERO population
is a key parameter in the study of the galaxy evolution
and can constrain models which link the formation of
massive elliptical galaxies and the onset of AGN activity (Granato et al. 2001;
Almaini et al. 2002). \\
Hard X-ray observations provide a
powerful tool to uncover AGN among the EROs population.
Indeed, a sizeable fraction of hard X-ray sources recently discovered
in deep \chandra\ and \xmm\ surveys are associated with EROs,
the exact value depending on the limiting fluxes reached in the
optical and X-ray bands (Brusa et al. 2002a; Mainieri et al. 2002;
Alexander et al. 2002). 
Moreover, the excellent imaging capabilities of \chandra\
enable to probe the average X--ray properties of EROs
even beyond the limiting flux of deep surveys, using the stacking
analysis technique.
%at the very deep limiting fluxes of Chandra 1 Ms exposures
%it is possible to further constrain their nature from the analysis 
%of their mean properties at high-energies.
\pn
In this framework we have developed two complementary programs of hard X-ray
observations of EROs making use of both \chandra\ and \xmm\ data.
First, we present the results of a \xmm\ survey on a large, complete
sample of EROs which is the most suitable one to estimate the fraction
of AGN--powered EROs among the optically selected population. 
Then, we have constrained the
high-energy properties of a sample of spectroscopically identified
non--AGN EROs in the \chandra\ Deep Field South (CDFS)
via stacking analysis.
\pn
Throughout the paper, a cosmology with $H_0=70$ km s$^{-1}$ Mpc$^{-1}$,
$\Omega_m$=0.3 and $\Omega_{\Lambda}$=0.7 is adopted. 

\section{\xmm\ observation of the ``Daddi Field''}
%####################################################
\subsection{X--ray data analysis and results}
%##########################################
We have started an extensive program of multiwavelength
observations of the largest sample of EROs available to date
($\sim$450 sources), selected in a contiguous area over a $\sim 700$
arcmin$^2$ field (the ``Daddi field'', Daddi et al. 2000) and
complete to a magnitude limit of K$=$19.2.
Deep optical (R$\sim 26.2$ at the 3$\sigma$ level) photometry is
available (see Daddi et al. 2000 for details in
optical and near--infrared data reduction) and VIMOS spectroscopy is 
planned at VLT.\\
The Daddi field was observed by \xmm\ on August 3, 2001 for a nominal exposure
time of $\sim 50$ ks.
The \xmm\ data were processed using version 5.2 of the Science Analysis System 
(SAS). The event files were cleaned up from hot pixels and soft proton
flares; % removing all the time intervals with a count rate higher
	% than 0.15 c/s in the 10--12.4 keV energy range for the two
	% MOS and higher than 0.35 c/s in the 10--13 keV band for the
	% {\it pn} unit. 
the resulting exposure time is $\sim 35$ ks in the MOS1 and MOS2
detectors and $\sim 30$ ks in the {\it pn} detector.
The {\em EBOXDETECT} task, the standard {\em SAS} sliding box cell
detect algorithm, was run on the 0.5--2, 2--8 and 0.5--8 keV cleaned events. 
We limited the X-ray analysis in a region of ten arcmin radius from
the centre of the XMM--{\it Newton} pointing, to take into account the
broadening of the \xmm\ PSF at increasing distance from the aim
point. Ninety--one sources were detected in the full X--ray band with a 
threshold in Poisson probability of 5$\times 10^{-6}$; 
the corrisponding flux limit is $\sim 3 \times 10^{-15}$ \cgs assuming
a power--law spectrum with $\Gamma$=1.8 and Galactic absorption
(N$_H=5\times10^{20}$ cm$^{-2}$).\\  
The X--ray centroids have been corrected for systematic errors with
respect to the optical positions of three bright quasars 
in the field (Hall et al. 2000) and then were cross--correlated with 
the $K$--band and $R$--band catalogs. 
\begin{figure}
\resizebox{\hsize}{!}
{\includegraphics[]{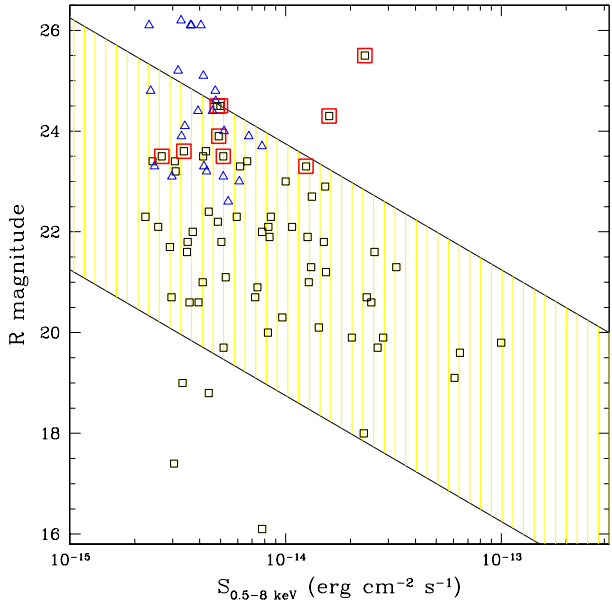}}
\caption{\small R--band magnitude vs. the 0.5--8 keV flux for the 91 sources
detected in the XMM--{\it Newton} observation. Squares are sources
detected both in the K--band (K$<19.2$) and in the R--band (R$<26.2$) 
images, triangles are sources detected in the R--band image only. 
Enlarged squares are EROs. The dashed area corresponds to the region of
$0.1<$f$_x$/f$_o <10$.}
\label{figlabel}
\end{figure}
We searched for near--infrared counterparts within a radius of
6$''$ from the X--ray position; we note, however, that the 75\% of the
sources lie within a 3$''$ radius circle. 
The results and the optical identifications are summarized in Fig. 1:
80\% of the 0.5--8 keV sources have at least one near--infrared
counterpart down to the K--band magnitude limit; four sources are not
detected at the R--band limit.
Nine out of ninety--one sources are associated with EROs ($\sim$10\%);
the fraction is the same (5/46) for the 2--8 keV sample.
We refer to Brusa et al. (2002a) for a more detailed discussion on the
\xmm\ data analysis, the associations of X--ray sources with optical
counterparts and the analysis of X--ray colors.

\subsection{AGN fraction in optically selected samples}
%#####################################################
The AGN fraction among optically selected EROs samples is still very
uncertain (Brusa et al. 2002a; Mainieri et al. 2002), and strongly
dependent on the limiting fluxes reached in both the X--ray and
near--infrared bands. 
About 370 EROs fall within the \xmm\ area analysed in this work.
The fraction of AGN--powered, hard X--ray selected  
EROs at the near--infrared (K=19.2) and X--ray flux limits of our
survey is therefore $\sim1\div$2\% (5/370).
This fraction rises up to 14\% at 
at K$\lsimeq$20.1 and S$_{2-10} \gsimeq 4\times 10^{-16}$ erg cm$^{-2}$
s$^{-1}$ limiting fluxes of the CDFN observation (Alexander et
al. 2002), supporting the idea that the bulk of the EROs population is
not related to active phenomena.  
\pn
It is worth noticing that these AGN--powered EROs, as well as 
others examples of hard X--ray selected EROs %(see Brusa et al. 2002a and reference therein) 
show a relatively well-defined correlation between the X--ray fluxes
and the optical magnitude around f$_{\rm X}$/f$_{\rm O}\simeq$ 10. 
This correlation is shifted by one order of magnitude from the one
found by {\tt ROSAT} for soft X--ray selected quasars (Hasinger et
al. 1998) and recently extended by {\it Chandra} and XMM--{\it Newton}
observations to hard X--ray selected sources (Alexander et
al. 2002; Lehmann et al. 2001).
The most plausible explanation of such a high X--ray--to--optical flux
ratio in these objects is the presence of high obscuration towards the
active nuclear source. This hypothesis is strongly supported by 
optical identifications available for a handful of objects (Cowie et
al. 2001; Mainieri et al. 2002; Hornschemeier et al. 2001) and by
the X--ray spectral analysis (see Sect. 4). 

\section{Stacking EROs in the CDFS}
%##########################################
The {\tt K20} EROs sample (Cimatti et al. 2002) %Daddi et al. 2002)
in the {\it Chandra} Deep Field South (CDFS) area
includes 48 objects at $K_s<20$.  
For the purposes of the present work we consider only the 
21 spectroscopically identified EROs which have been 
classified in two categories:
13  ``dusty'', i.e. objects showing at least an emission line (tipically
[OII]$\lambda3727$) over a red continuum, and 8 ``old'', i.e. objects
with an absorption--line spectrum, consistent with that of early--type,
passively evolving galaxies, distributed on a similar 
range of redshifts ($z=0.8\div1.6$).\\
For the X--ray analysis, we made use of the public CDFS
one megasecond {\it Chandra} observation available in the archive
(Giacconi et al. 2002).  
At first, we have searched for individual X--ray emission from
all the objects in the EROs spectroscopic redshift catalog of Cimatti
et~al. (2002).
Only one source, a dusty ERO at $z=1.327$, was 
individually detected in the one megasecond exposure. 
The optical position is almost coincident ($\Delta <0.3"$, within the
accuracy of the K--band position) %Cimatti et al. 2002b) 
with the {\it Chandra} source CXO CDFS~J033213.9-274526
in the Giacconi et al. (2002) catalog.
The very hard X--ray spectrum and the high intrinsic X--ray luminosity
unambiguously reveal the presence of an obscured AGN in this source.
Indeed, this object is undetected in the soft band, implying an
intrinsic column density larger than 4$\times10^{23}$ cm$^{-2}$, 
and an unabsorbed hard X--ray luminosity L$_{2-10}>4\times10^{43}$ 
erg s$^{-1}$; its properties are then more extreme than those 
inferred from the 2 Ms CDF--N spectral analysis (Vignali et al. 2002). 
Given the small area sampled and the spectroscopic incompleteness,
the present ERO sample is not suitable for a reliable estimate of
the AGN fraction among the EROs population. However, the present
low detection rate of AGN among EROs (1/21) confirms that the bulk
of the EROs population is not related to active phenomena (see
Sect. 2.2).\\
In order to constrain the average X--ray properties of individually 
undetected EROs, we have applied the 
stacking technique (Brandt et al. 2001; Nandra et al. 2002)  
separately for the two classes of objects. The samples 
consist of 12 ``dusty'' and 8 ``passive'' EROs,  
with average redshifts of $z=1.053$  and $z=1.145$, respectively.
We performed the detection in the three standard X--ray bands (0.5--8,
0.5--2 and 2--8 keV) and in the 1--5 keV band which roughly
corresponds to the rest--frame hard X--ray band at the average
redshift of the samples.
We refer to Brusa et al. (2002b) for details on the X--ray
data analysis.\\
In Fig.~2 we show the summed images in the full (0.5--8 keV) band.
The most important finding is that the two spectroscopically classes of EROs
have different X--ray properties. 
\begin{figure}
\resizebox{\hsize}{!}
{\includegraphics[]{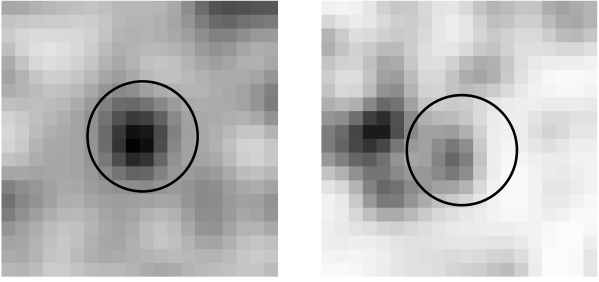}}
\caption{\small Stacked full--band images of the ``dusty'' ({\it left
panel}) and ``old'' ({\it right panel}) EROs.
The images are 9x9 arcsec and have been smoothed with a gaussian of
$\sigma$=1.5 pixel (approx 0.7$''$) .
The circles are centered on the stacking position and have
a radius of 2$''$. The detection significance of the summed counts is
4.2$\sigma$ for the ``dusty'' and $<2\sigma$ for the ``old''.
The bright spot in the right panel at about 4$''$ from the stacked
position, corresponds to an individually detected object.}
\label{figlabel}
\end{figure}
%\pn
\subsection{An X--ray dichotomy}
%#################################
The X--ray emission from ``old'' EROs remains undetected 
in all the considered energy bands, for a total effective exposure 
time of $\sim 7.0$ Ms. Assuming a thermal emission model with kT=1 keV
and solar metallicity, fully consistent with the average X--ray
spectrum of nearby elliptical galaxies (e.g., Pellegrini 1999), the
3$\sigma$ upper limit on the 0.5--2 keV luminosity is $L_X < 10^{41}$
erg s$^{-1}$. 
\pn
An excess of counts above the expected background level is clearly 
present in the ``dusty'' sample, for a total effective exposure time
of 10 Ms. The signal is stronger in the full band (99.997\% confidence level
assuming a Poisson distribution, corresponding to about
4.2$\sigma$) and it is still present,  
although at a slightly lower confidence level, in the soft and 
1--5 keV band, while it is not statistically significant in the hard
band.
Assuming a $\Gamma = 2$ power--law spectrum plus Galactic absorption 
(N$_H=8\times 10^{19}$ cm$^{-2}$), 
the 3$\sigma$ upper limit on the average band ratio 
(H/S$<$0.85, where H and S indicates the hard band 
and soft band counts, respectively),  
corresponds to an absorption column density lower than $10^{22}$ cm$^{-2}$
(see Fig. 4 in Alexander et al. 2002). 
Given that the hard X--ray spectrum of nearby starburst galaxies is
well--represented by power--law models with slopes in the range
1.8$\div 2.5$ and low intrinsic absorption (e.g., Ptak et
al. 1999; Ranalli et al. 2002), we adopted $\Gamma$=2.1 to compute
X--ray fluxes and luminosities. 
With this assumption, the stacked count rate in the 1--5 keV band
corresponds to an average 2--10 keV rest--frame luminosity of  
$\sim 8 \times 10^{40}$~erg~s$^{-1}$, at the mean redshift $z=1.053$. 
The inferred, average X--ray luminosity 
is around one order of magnitude larger than that of normal spiral 
galaxies (Matsumoto et al. 1997), it is similar to that of the 
starburst galaxy M82 (e.g., Griffiths et~al. 2000; Kaaret et~al. 2001)
and about 3 times lower than that of NGC~3256, one of the most 
X-ray luminous starburst in the local Universe 
(e.g., Moran, Lehnert, \& Helfand 1999). \\
The starburst scenario is supported also by the optical data.
Indeed, following Cimatti et al. (2002), we recomputed the 
average spectrum of the 12 EROs we used in the X--ray stacking analysis.
The average optical spectrum shows a very red and smooth continuum, 
with a strong [OII]$\lambda3727$ emission and lacks of the presence 
of clear AGN indicators such as [NeV]$\lambda3426$,
suggesting that the optical emission is dominated by star--forming
systems. 

\section{Hard X--ray properties of EROs}
%#######################################
We have collected from the literature a sample of hard X--ray selected EROs
serendipitously detected in moderate--deep and deep \chandra\ and \xmm\
observations (Hornschemeier et al. 2001; Alexander et al. 2002; 
Mainieri et al. 2002). From the observed X--ray band ratio, we have
computed for each ERO the corresponding X--ray column density for a
source with a typical AGN power--law spectrum ($\Gamma$=1.8) at $z=1$,
while for the sources in the Lockman Hole we adopted the best--fit
value quoted by Mainieri et al (2002) and corrected to $z=1$ for
unidentified sources. The results are reported in Fig. 3.
Almost all of the individually detected sources are consistent with
intrinsic column densities in excess of 10$^{22}$
cm$^{-2}$. 
The so far individually detected EROs
have high X--ray luminosity (L$_X$$>$10$^{43}$ erg s$^{-1}$, in a few
case even larger than $10^{44}$ erg s$^{-1}$) and they actually are
heavily obscured AGN, as inferred from X--ray spectral analysis  
(Cowie et al. 2001; Gandhi et al. 2002; Mainieri et al. 2002). 
Therefore, hard X--ray selected EROs (or at least a fraction of them) 
have properties similar to those of Quasars 2, the high--luminosity, 
high redshift type II AGNs predicted in X--Ray Background synthesis
models (e.g. Comastri et al. 1995). 
\pn
Stacking analysis of those objects not individually detected in the
X--rays suggests a softer X--ray spectrum than that of AGN--powered
EROs (see also Alexander et al. 2002).
Moreover, the results from the CDFS {\tt K20} EROs sample suggest that
different spectroscopic classes of EROs are characterized by different
high--energies properties. 
Indeed, ``dusty'' EROs are moderately luminous
hard X--ray emitters and their average X--ray spectrum (N$_H<10^{22}$
cm$^{-2}$) is consistent with that measured for nearby star--forming
galaxies (e.g. Dahlem, Weaver \& Heckman 1998).
Early--type galaxies among EROs are not detected in the X--rays
and their 3$\sigma$ upper limit on the soft X--ray luminosity is
consistent with the emission from nearby elliptical galaxies
(Pellegrini 1999).

\begin{figure}
\resizebox{\hsize}{!}{
\includegraphics[]{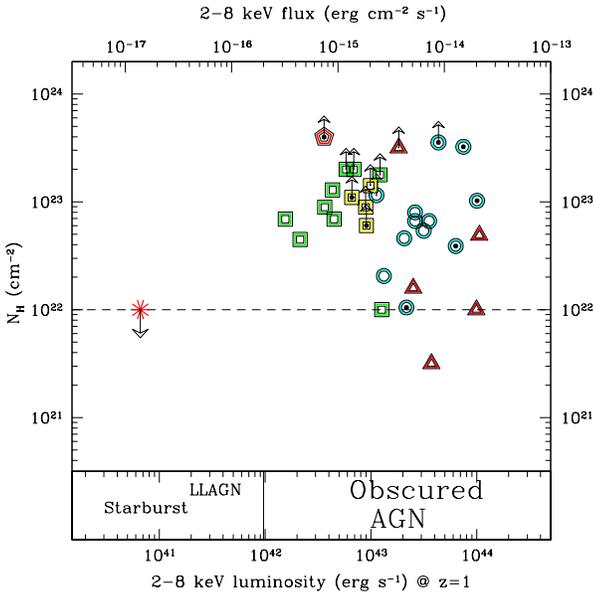}}
\caption{\small X--ray absorption column density vs. hard X--ray flux for a
sample of AGN--powered EROs, serendipitously detected in hard X--ray
surveys. Different symbols refer to different samples as follows: 
triangles = ``Daddi'' field; squares = CDFN (Hornschemeier et
al. 2001; Alexander et al. 2002); circles =  Lockman Hole (Mainieri et
al. 2002). 
The individually detected AGN in the {\tt K20} CDFS area is marked
with a pentagon; the result from the stacking analysis of the
``dusty'' sample is marked with an asterisk. 
Dot--filled symbols refer to sources with measured redshift.}
\label{figlabel}
\end{figure}
\section{Conclusions}
%###########################
The most important results can be summarized as follows:
\begin{itemize}
\item[$\bullet$] Hard X--ray observations turned out to be extremely
useful to pick--up AGN--powered EROs in optically selected samples.
The fraction of AGN among EROs is in the range $1\div15$\%, depending
on the limiting fluxes in the X--ray and near--infrared bands.
\item[$\bullet$] AGN--powered EROs are X--ray luminous and
obscured objects. The intrinsic column densities inferred for a sample
of hard X--ray selected EROs are in excess of $10^{22}$ cm$^{-2}$.
\item[$\bullet$] ``Dusty'' EROs are relatively unobscured 
(N$_H<10^{22}$ cm$^{-2}$) hard X--ray sources and their X--ray
properties are consistent with those of nearby star--forming galaxies.
\item[$\bullet$] ``Old'' EROs are not detected in the X--rays.
\end{itemize}

\acknowledgements
I thank A. Comastri, A. Cimatti, E. Daddi, M. Mignoli, L. Pozzetti and
C. Vignali for fruitful discussion and for a careful reading of the
manuscript.
This work was supported by ASI I/R/113/01
and MIUR Cofin--00--02--36 grants.

\end{document}